\documentclass[twoside,12pt]{article}
\usepackage{epsfig}

\def\p{\vec p}

\def\sgr0526{\mbox{SGR~0526-66}}

\def\groj1744{\mbox{GRO~J1744-28}}
\def\rxj1856{\mbox{RX~J1856.5-3754}}
\def\0720{\mbox{RX~J0720.4-3125}}
\def\eu1728{\mbox{4U~1728-34}}
\def\3c58{\mbox{3C58}}
\def\1e1207{\mbox{1E~1207.4-5209}}
\def\p0943{\mbox{PSR~0943+10}}

\def\ls{\lower0.5ex\hbox{$\; \buildrel < \over \sim \;$}}
\def\gs{\lower0.5ex\hbox{$\; \buildrel > \over \sim \;$}}

\def\slash#1{#1\!\!\!/}

\newcommand{\be}{\begin{equation}}
\newcommand{\ee}{\end{equation}}
\newcommand{\bea}{\begin{eqnarray}}
\newcommand{\eea}{\end{eqnarray}}

\newcommand{\msun}{{M_{\odot}}}

\newcommand{\mevt}{{\rm MeV/fm}^3}
\newcommand{\gcmt}{{\rm g/cm}^3}

\newcommand{\btau}{\mbox{\mbox{\boldmath$\tau$}}}

\newcommand{\bPhi}{\mbox{\mbox{\boldmath$\Phi$}}}

\newcommand{\bfAm}{\mbox{\mbox{\boldmath$A$}}}

\newcommand{\ergs}{{\rm erg/s}}
\newcommand{\mev}{{\rm MeV}}

\newcommand{\kFo}{k_{F_0}}
\newcommand{\kFosq}{{k_{F_0}^2}}

\newcommand{\nuk}{{\nu_{\rm K}}}
\newcommand{\psib}{\bar{\psi}}
\newcommand{\Lcal}{{\mathcal L}}

\begin{document}

\title{\vspace{1cm} Pulsars as Astrophysical Laboratories for Nuclear
\\ and Particle Physics}

\author{F.\ Weber,$^{1}$ R.\ Negreiros,$^{1}$ P.\ Rosenfield,$^{1}$ \\
M.\ Stejner$^{1,2}$\\ \\ $^1$Department of Physics, San
Diego State, University, \\ 5500 Campanile Drive, San Diego,
California 92182, USA\\ $^2$Department of Physics and Astronomy,
University of Aarhus, \\ Ny Munkegade, DK-8000 Aarhus C, Denmark}

\maketitle

\begin{abstract} 
A forefront area of research concerns the exploration of the
properties of hadronic matter under extreme conditions of temperature
and density, and the determination of the equation of state--the
relation between pressure, temperature and density--of such matter.
Experimentally, relativistic heavy-ion collision experiments enable
physicists to cast a brief glance at hot and ultra-dense matter for
times as little as about $10^{-22}$~seconds.  Complementary to this,
the matter that exists in the cores of neutron stars, observed as
radio pulsars, X-ray pulsars, and magnetars, is at low temperatures
but compressed permanently to ultra-high densities that may be more
than an order of magnitude higher than the density of atomic
nuclei. This makes pulsars superb astrophysical laboratories for
medium and high-energy nuclear physics, as discussed in this paper.
\end{abstract}





\goodbreak
\section{Introduction}\label{sec:intro}

\begin{figure}[tb]
\begin{center}
\centerline{\psfig{figure=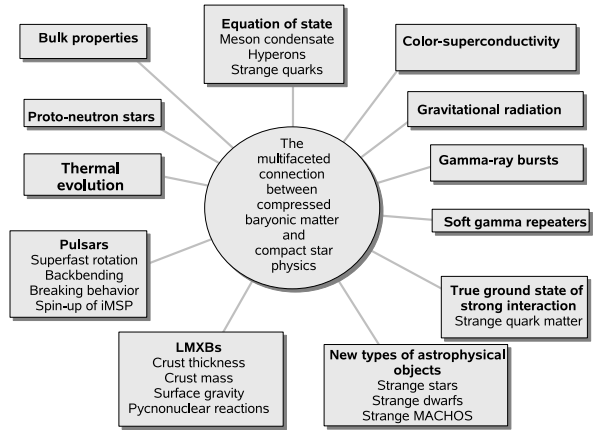,width=8.0cm}}
\begin{minipage}[t]{16.5 cm}
\caption[]{The multifaceted connection between compressed baryonic
 matter matter and pulsar physics \cite{weber05:a}.}
\label{fig:multifaceted2} 
\end{minipage}
\end{center}
\end{figure}

Exploring the properties of hadronic matter under extreme conditions
of temperature and density has become a forefront area of modern
physics, both theoretically and experimentally. On the earth, heavy
ion experiments enable physicists to cast a brief glance at such
matter for times as little as about $10^{-22}$~seconds.  On the other
hand, it is well known that galaxies like our Milky Way contain up to
$10^8$ to $10^9$ neutron stars, which are observed as pulsars
(rotating neutron stars). Such objects contain ultra-dense hadronic
matter as a permanent component of matter in their centers.  Radio
telescopes, X-ray satellites--and soon the latest generation of
gravitational-wave detectors--provide physicists and astronomers with
an unprecedented wealth of high-quality data on such objects and thus
serve as the observational windows on the inner workings of pulsars.
This feature makes pulsars superb astrophysical laboratories for
medium and high-energy nuclear physics (plus other fields of physics)
\cite{glen97:book,weber99:book,heiselberg00:a,lattimer01:a,blaschke01:trento,%
weber05:a,sedrakian06:a,klahn06:a_short}.  Some of the key questions
that can be addressed by studying the properties of pulsars are (see
Fig.\ \ref{fig:multifaceted2}):
\begin{itemize}
\item What are the fundamental building blocks of cold ultra-dense
matter?  Specifically, does ``exotic'' matter exist in the cores of
pulsars, such as boson condensates, color-superconducting quark
matter, and multi-quark states?

\item Are there pulsar observables that could signal the existence of
exotica in their cores?

\item Are there rotationally-driven (accretion-driven) phase
transitions in pulsars?

\item How does color-superconducting quark matter alter the properties
of pulsars?

\item What is the true ground state of the strong interaction? Is it
ordinary nuclear matter (i.e.\ atomic nuclei) or a color-neutral
collection of up, down, and strange quarks (so-called strange quark
matter)?

\item What are the distinguishing features of pulsars made of
strange quark matter rather than ordinary hadronic matter? 

\item What are the properties of matter subjected to ultra-high
density radiation fields, ultra-high magnetic fields, ultra-high
electric fields? How do such fields alter the properties of pulsars?

\item What are the key nuclear (heavy ion) reactions in the
non-equilibrium crusts of accreting X-ray pulsars? 

\item How strongly do pycnonuclear reactions in the crusts 
of accreting neutron stars alter the thermal evolution of such objects?

\item Do gravitational-radiation reaction driven instabilities limit the 
spins of pulsars?

\item What is the shell structure for very neutron rich nuclei in the 
crusts of pulsars? Do N=50 and N=82 remain magic numbers? 

\end{itemize}
Before we discuss some of these issues in more detail, we review
some of the key properties of neutron stars.  The first property
concerns the masses of neutron stars, which range theoretically from
around $0.1\,\msun$ (where $\msun=2\times 10^{33}$~g is the mass of
the sun) to about $3\,\msun$.  Matter in the
centers of neutron stars possess densities ranging from a few times
$n_0$ to an order of magnitude higher.  Here $n_0=0.15$
nucleons/fm$^3$ denotes the baryon number density of normal nuclear
matter, which corresponds to a mass density of $2.5\times
10^{14}$~g/cm$^3$. The number of baryons forming a neutron star is of
the order of $10^{57}$.  Rotating neutron stars are called
pulsars. Three distinct classes of pulsars are currently known. These
are (1) rotation-powered pulsars, where the loss of rotational energy
of the star powers the emitted electromagnetic radiation, (2)
accretion-powered (X-ray) pulsars, where the gravitational potential
energy of the matter accreted from a low-mass companion is the energy
source, and (3) magnetars (e.g, SGR 1806-20), where the decay of a
ultra-strong magnetic field powers the radiation.  The fastest, very
recently discovered neutron star, PSR J1748-2446ad, rotates at a
period of 1.39~ms (which corresponds to 719~Hz) \cite{hessels06:a},
followed by PSR B1937+21 \cite{backer82:a} and PSR B1957+20
\cite{fruchter88:a} whose rotational periods are 1.58 ms (633~Hz) and
1.61~ms (621~Hz), respectively.  PSR B1957+20 is moving through the
galaxy at a speed of almost a million kilometers per hour. Due to this
remarkable motion a bow shock wave is visible to optical telescopes.
Other significant pulsars are Cen X-3 (first X-ray pulsar), SAX
J1808.4-3658 (first accreting millisecond X-ray pulsar), and PSR
J0737-3039A{\&}B (first double pulsar binary system, which will permit
a strong-field test of General Relativity).  When neutron stars are
formed they have interior temperatures of the order of $10^{11}$~K
(around $10$~MeV). They cool by neutrino emission processes to
interior temperatures of $\sim 10^{10}$~K within a few
days. Throughout most of the active life of neutron stars in pulsars
and X-ray sources, the interior temperature is $\sim 10^7 - 10^9$ K.
Surface temperatures are an order of magnitude or more smaller.
Measurements of the surface temperature of the Crab pulsar (B0531+21),
for instance, have led to $T < 2 \times 10^6$~K.)  Rotating magnetized
neutron stars in pulsars are surrounded by a plasma, the so-called
magnetosphere, in which via plasma processes the electromagnetic
radiation from pulsars is generated. The magnetic fields near the
surface of neutron stars in compact X-ray sources play an important
role in channeling the accreting matter onto the neutron star surface.
\begin{figure}[tb]
\begin{center}
\centerline{\psfig{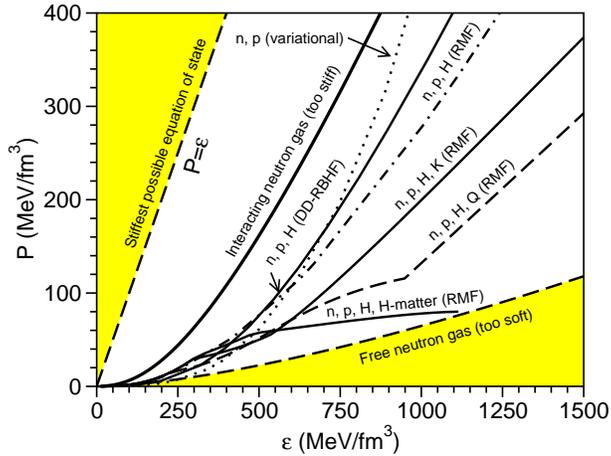}}
\begin{minipage}[t]{16.5 cm}
\caption[]{Models for the EoS (pressure versus energy 
density) of neutron star matter \cite{weber05:a}. The notation is as
follows: RMF=relativistic mean-field model; DD-RBHF=density dependent
relativistic Brueckner-Hartree-Fock model; n=neutrons; p=protons;
H=hyperons, K=$K^-[u,\bar s]$ meson condensate; Q=$u,d,s$ quarks;
H-matter=H-dibaryon condensate.}
\label{fig:nucl_eoss} 
\end{minipage}
\end{center}
\end{figure} 
To date over 1600 pulsars are known. This figure is expected to
increase dramatically with the operation of instruments like the
Square-Kilometer-Array (SkA). The sensitivity of SkA will be around
100 times higher than the VLA sensitivity; it is expected that around
20,000 new pulsars (including pulsars around black holes) will be
discovered with SkA. The number of millisecond pulsars is expected to
go up from its present value by a factor of 100. The initial
operations of SkA will start around 2016, and the final operations are
expected to begin around 2020.

\section{Bounds on the Nuclear Equation of State}\label{sec:masses}

In 1939, Tolman, Oppenheimer and Volkoff performed the first neutron
star calculations, assuming that such objects are entirely made of a
gas of non-interacting relativistic neutrons
\cite{oppenheimer39,tolman39:a}. The EoS of such a gas is extremely
soft, i.e.\ very little additional pressure is gained with increasing
density, as can be seen from Fig.\ \ref{fig:nucl_eoss}, and predicts a
\begin{figure}[tb]
\begin{center}
\centerline{\psfig{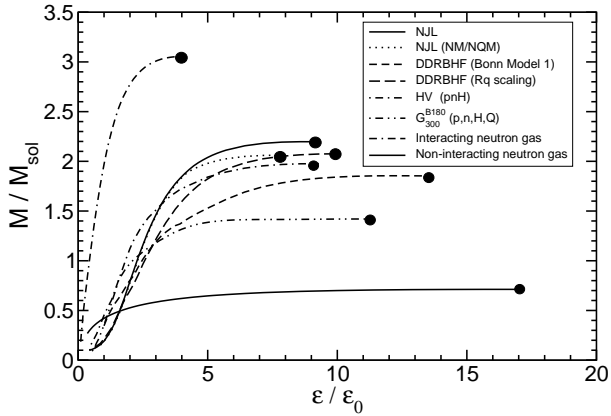}}
\begin{minipage}[t]{16.5 cm}
\caption{Neutron star mass versus central energy density ($\epsilon_0
= 140~\mevt$).}
\label{fig:ec} 
\end{minipage}
\end{center}
\end{figure} 
maximum neutron star mass of just $0.7~\msun$ (Figs.\ \ref{fig:ec} and
\ref{fig:MvsR}) at an unrealistically high density of 17 times the
density of nuclear matter (Fig.\ \ref{fig:ec}).  It is interesting to
note that the inclusion of interactions among the neutrons increases
the star's maximum mass from $0.7~\msun$ to around $3~ \msun$ (Figs.\
\ref{fig:ec} and \ref{fig:MvsR}). However, the radii of the latter
stars are so big that mass shedding from the equator occurs for most
stars of the sequence at rotational frequencies that are considerably
smaller than 716~Hz observed for PSR~J1748-2446ad, or B1937+21,
630~Hz (1.58~ms) \cite{backer82:a}. An interacting neutron gas thus
fails to accommodate the rotational frequencies of the most rapidly
rotating, observed neutron stars.  The other extreme, a
non-interacting relativistic neutron gas, fails too since it does not
accommodate the Hulse-Taylor pulsar ($M=1.44\, \msun$)
\cite{taylor89:a}, and also conflicts with the average neutron star
mass of $1.350 \pm 0.004\, \msun$ derived by Thorsett and Chakrabarty
\cite{thorsett99:a} from observations of radio pulsar systems.  More
than that, recent observations indicate that neutron star masses may
be as high as around $2~ \msun$. Examples of such very heavy neutron
stars are $M_{\rm J0751+1807} = 2.1 \pm 0.2 ~ \msun$ \cite{nice05:b},
$M_{\rm 4U\,1636+536} = 2.0 \pm 0.1~ \msun$ \cite{barret06:a}, $M_{\rm
Vela\, X-1} = 1.86\pm 0.16\, \msun$ \cite{barziv01:a}, $M_{\rm Cyg\,
X-2} = 1.78\pm 0.23 \, \msun$ \cite{casares98:a,orosz99:a}.  Large
masses have also been reported for the high-mass X-ray binary
4U\,1700--37 and the compact object in the low-mass X-ray binary
2S0921--630, $M_{\rm 4U\,1700-37} = 2.44 \pm 0.27~ \msun$
\cite{clark02:a} and $M_{\rm 2S0921-630} = 2.0 - 4.3 \msun$
\cite{shahbaz04:a}, respectively. The latter two objects may be either
massive neutron stars or low-mass black holes with masses slightly
higher than the maximum possible neutron star mass of $~\sim 3
\msun$. This value follows from a general, theoretical estimate of the
maximal possible mass of a stable neutron star \cite{rhoades74:a}.  If
either one of the two objects 4U\,1700--37 or 2S0921--630 were a black
hole, it would confirm the prediction of the existence of low-mass
black holes \cite{brown94:a}. Conversely, if these objects were
massive neutron stars, their high masses would severely constrain the
EoS of dense nuclear matter.

\section{Nuclear Many-Body Models}

A vast number of models for the equation of state (EoS) of neutron
star matter has been derived in the literature over the years. The
majority of these models belong to either one of the following
categories: \\
{\bf 1.} Thomas-Fermi based models \cite{myers95:a,strobel97:a} such as the
new Thomas-Fermi approach of Myers and Swiatecki \cite{myers95:a}. It
is based on a Seyler-Blanchard potential generalized by the addition
of one momentum dependent and one density dependent term
\begin{eqnarray}
  V_{12} = - \frac{2\, T_{0}}{\rho_0} \, Y\bigl(r_{12}\bigr) \left(
 \frac{1}{2} (1 \mp \xi) \alpha - \frac{1}{2}(1 \mp \zeta) \Bigl(\beta
 \Bigl( \frac{p_{12}}{\kFo} \Bigr)^2 - \gamma \frac{\kFo}{p_{12}} +
 \sigma \Bigl( \frac{2 \bar \rho}{\rho_0} \Bigr)^{\frac{2}{3}} \Bigr)
 \right) \, .
\label{eq:v.tf96}
\end{eqnarray} The upper (lower) sign in Eq.\ (\ref{eq:v.tf96}) corresponds
to nucleons with equal (unequal) isospin.  The quantities $\kFo$,
$T_{0}$ $(= \kFosq / 2m)$, and $\rho_0$ are the Fermi momentum, the
Fermi energy and the saturation density of symmetric nuclear
matter. The potential's radial dependence is described by a
Yukawa-type interaction of the form $Y (r_{12}) = (4 \pi a^3)^{-1}
\exp(-r_{12}/a)/ (r_{12}/a)$. Its strength depends both on the
magnitude of the particles' relative momentum, $p_{12}$, and on an
average of the densities at the locations of the particles.  The
parameters $\xi$ and $\zeta$ were introduced in order to achieve
better agreement with asymmetric nuclear systems. The behavior of the
optical potential is improved by the term $\sigma
(2\bar\rho/\rho_0)^{2/3}$ with the average density defined as
$\overline{\rho}^{2/3} = (\rho_{1}^{2/3} + \rho_{2}^{2/3})/2$, where
$\rho_1$ and $\rho_2$ the densities of interacting neutron or protons
at points 1 and 2. The seven free parameters of the theory are
adjusted to the properties of finite nuclei, the parameters of the
mass formula, and the behavior of the optical potential
\cite{myers95:a}. \\ 
{\bf 2.} Schroedinger-based models
\cite{heiselberg00:a,sedrakian06:a,pandharipande79:a,wiringa88:a,akmal98:a}
are derived from Hamiltonians of the form
\begin{equation}
  {\cal H} = \sum_i \, {{-\,\hbar^2}\over{2\, m}} \, \nabla_i^2 +
  \sum_{i<j} \, {V}_{ij} + \sum_{i<j<k} \, {V}_{ijk} \, ,
\label{eq:hamil}
\end{equation}
where $V_{ij}$ and $V_{ijk}$ denote two and three-nucleon
interactions. The many-body equations are then solved, for instance,
in the framework of the hole-line expansion (Brueckner theory),
coupled cluster method, self-consistent Green functions technique, or
variational approach. \\
{\bf 3}. Relativistic nucler field-theoretical treatments such as
relativistic mean field (RMF), Hartree-Fock (RHF), standard
Brueckner-Hartree-Fock (RBHF), density dependent RBHF (DD-RBHF)
\cite{glen97:book,weber99:book,lenske95:a,fuchs95:a,typel99:a,hofmann01:a,%
niksic02:a,ban04:a} which are based on a Lagrangian of the form $\Lcal
= \Lcal_{B} + \Lcal_{M} + \Lcal_{int} + \Lcal_{lept}$, where 
\begin{eqnarray}
\Lcal_{B} &=& \psib \left( i\gamma_\mu\partial^\mu - m \right) \psi \, ,
\label{eq:LB} \\ 
\Lcal_{M} &=&\frac{1}{2} \sum_{i=\sigma,\delta}
\left(\partial_\mu\Phi_i\partial^\mu\Phi_i - m_i^2\Phi_i^2\right) -
\frac{1}{2} \sum_{\kappa=\omega,\rho} \left( \frac{1}{2}
F^{(\kappa)}_{\mu\nu} F^{(\kappa)\mu\nu} - m_\kappa^2 A^{(\kappa)}_\mu
A^{(\kappa)\mu} \right) \, , \label{eq:Lagrangian} \\
\Lcal_{int} &=&\psib\hat{\Gamma}_{\sigma}(\psib,\psi)\psi\Phi_{\sigma}
- \psib\hat{\Gamma}_{\omega}(\psib,\psi)\gamma_{\mu}\psi A^{(\omega)
  \mu} + \psib\hat{\Gamma}_{\delta}(\psib,\psi) 
{\btau} \psi {\bPhi}_{\delta} - \psib\hat{\Gamma}_{\rho}
(\psib,\psi)\gamma_{\mu} {\btau} \psi {\bfAm}^{(\rho)\mu} \,
. \label{eq:Lint}
\end{eqnarray}
Here, $\Lcal_B$ and $\Lcal_M$ are the free baryonic and the free
mesonic Lagrangians, respectively, and interactions are described by
$\Lcal_{int}$, where $ F^{(\kappa)}_{\mu\nu} = \partial_\mu
A_\nu^{(\kappa)} - \partial_\nu A_\mu^{(\kappa)}$ is the field
strength tensor of one of  the vector mesons ($\kappa= \omega, \rho$).
In RMF, RHF and RBHF the meson-baryon vertices $\hat\Gamma_\alpha$
($\alpha=\sigma,\omega,\delta, \rho$) are density-independent
quantities which are given by expressions like $\hat\Gamma_\sigma = i
g_\sigma$ for the scalar $\sigma$ meson, $\hat\Gamma^\mu_\omega =
g_\omega \gamma^\mu + (i/2) (f_\omega / 2 m) \partial_\lambda
[\gamma^\lambda,\gamma^\mu]$ for $\omega$ mesons,
etc. \cite{weber99:book}. In the framework of the DD-RBHF scheme, 
 the meson-baryon vertices $\hat\Gamma_\alpha$ are not only determined by
Dirac matrices but depend on the baryon field operators $\psi$
\cite{hofmann01:a}. \\
{\bf 4.} The Nambu-Jona-Lasinio (NJL) model was originally introduced to
describe nucleons with dynamically generated masses. In recent years
it has become a popular model to describe quarks--hadron matter, and
to explore the condensation patter of color superconducting quark
matter
\cite{alford98:a,rapp98:a,buballa05:a,blaschke05:a,rischke05:a,abuki06:a,%
lawley06:a,lawley06:b}. The flavor SU(2) NJL-Lagrangian, for instance,
which has been applied in \cite{lawley06:b} to derive both nuclear
matter and quark matter phases, is given by
\begin{eqnarray}
{\cal L} &=& \bar{\psi} (i \slash\partial - m_q) \psi +
G_{\pi} \left( (\bar{\psi}\psi)^2 - (\bar{\psi}\gamma_5
{\btau}\psi)^2\right)  
- G_{\omega} (\bar{\psi}\gamma^{\mu}\psi)^2
- G_{\rho} (\bar{\psi}\gamma^{\mu}{\btau}\psi)^2 \nonumber \\
&&+ G_s (\bar{\psi}\gamma_5 C \tau_2 \beta^A \bar{\psi}^T)
(\psi^T C^{-1} \gamma_5 \tau_2 \beta^A \psi) \, .
\label{lag}
\end{eqnarray}
Here $m_q$ is the current quark mass, $\psi$ is the flavor SU(2) quark
field, and the coupling constants $G_{\pi}$, $G_{\omega}$ and
$G_{\rho}$ characterize the $q{\bar q}$ interactions in the scalar,
pseudoscalar and vector meson channels, while $G_s$ refers to the
interaction in the scalar diquark channel \cite{buballa05:a,lawley06:b}.\\
{\bf 5.} Aside from the NJL model, there are several other
phenomenological models based on quark degrees of freedom, such as the
quark meson coupling model, the cloudy bag model, the quark mean field
model, and the chiral SU(3) quark mean field model \cite{wang05:a}.
In the latter model, quarks are confined within baryons by an
effective potential. The quark-meson interaction and meson
self-interaction are based on SU(3) chiral symmetry. The chiral SU(3)
quark mean field model was applied  recently to the study of neutron stars 
and strange stars \cite{wang05:a}. 

\begin{figure}[tb]
\begin{center}
\parbox[t]{7.5cm}
{\psfig{figure=MvsR.eps,width=7.0cm}
{\caption[]{Mass--radius relationship of neutron stars and strange
stars \cite{weber05:a}.}
\label{fig:MvsR}}}
\ \hskip 0.1cm \
\parbox[t]{7.5cm} 
{\psfig{file=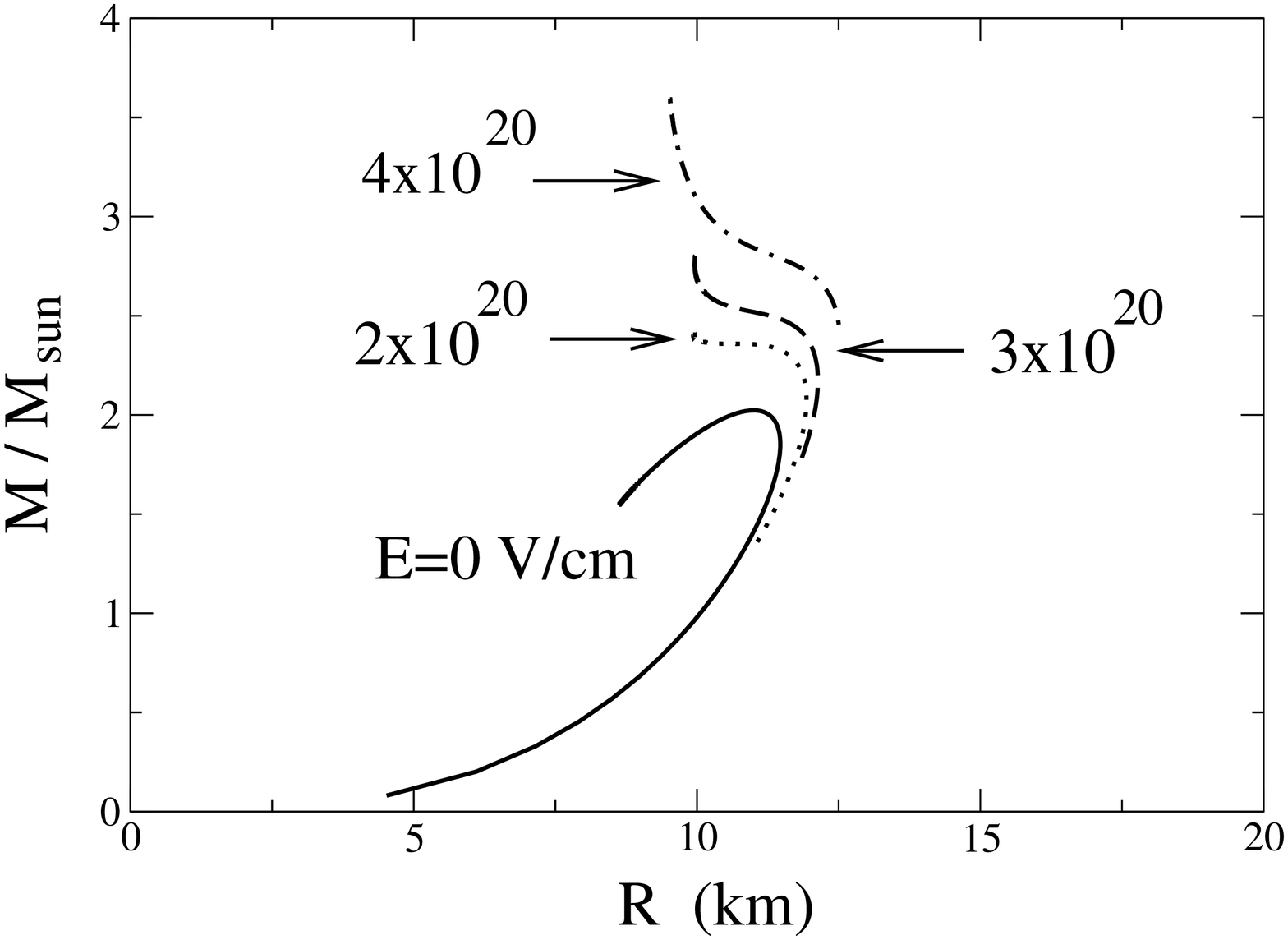,width=7.0cm,angle=0}
{\caption[]{Mass--radius relationship of electrically charged strange stars.}
\label{fig:MvsR.charged}}}
\end{center}
\end{figure}

A collection of equations of state computed for several of these
models is shown in Fig.\ \ref{fig:nucl_eoss}. Mass--radius
relationships of neutron stars based on these EoS are shown in Figs.\
\ref{fig:MvsR} and \ref{fig:MvsR.charged}. Strange star sequences are
shown in these figures too.
The strange star sequences shown in Fig.\ \ref{fig:MvsR.charged} show
the impact of ultra-strong electric fields of the mass--radius
relationship of strange stars. The energy-momentum tensor of such
stars consists of the usual perfect-fluid term, which is supplemented
with the electromagnetic energy-momentum tensor \cite{ray03:a}, $
T_{\nu}{}^{\mu} = (p +\rho c^2)u_{\nu} u^{\mu} + p\delta_{\nu}{}^{
\mu} + [ F^{\mu l} F_{\nu l} + \delta_{\nu}{}^{\mu} F_{kl} F^{kl} / 4
\pi ] / 4 \pi $,
where $u^{\mu}$ is the fluid's four-velocity, $p$ and $\rho c^2\equiv
\epsilon$ are the pressure and energy density, respectively, and
$F^{\mu \nu}$ satisfies the covariant Maxwell equation, $[(-g)^{1/2}
F^{\nu \mu}]_{, \mu} = 4\pi J^{\nu} (-g)^{1/2}$. The total mass of the
star, contained within a radial distance $r$ from the star's center,
is given by $ dm(r)/dr = (4\pi \epsilon r^2)/ c^2 + (Q(r)/ c^2 r)
(dQ(r)/dr)$. This relation shows that, in addition to the standard term
originating from the EoS of the stellar fluid, the electric field
energy too contributes to the star's total mass.  The
Tolman-Oppenheimer-Volkoff (TOV) equation of electrically charged
stars is given by
\begin{eqnarray}
\frac{dp}{dr} = - \frac{2G\left[ m(r) +\frac{4\pi r^3}{c^2} \left( p -
\frac{Q^{2} (r)}{4\pi r^{4} c^{2}} \right) \right]}{c^{2} r^{2} \left(
1 - \frac{2Gm(r)}{c^{2} r} + \frac{G Q^{2}(r)}{r^{2} c^{4}} \right)}
(p +\epsilon) +\frac{Q(r)}{4 \pi r^4}\frac{dQ(r)}{dr} \,
. \label{TOVca}
\end{eqnarray}
As shown in \cite{ray03:a}, electric fields can substantially alter
the structure of compact stars. This is specifically the case for the
masses of neutron stars, provided they posses net electric charges. For
strange stars, however, even the maximum possible electric fields,
$\sim 10^{18}$~V/cm, modifies the mass--radius relationship only very
weakly as can be seen from Fig.\ \ref{fig:MvsR.charged}.

\goodbreak
\section{Building Blocks of Neutron Star Matter}

\subsection{\it Hyperons and baryon resonances}

\begin{figure}[tb]
\begin{center}
\parbox[t]{7.55cm} {\psfig{file=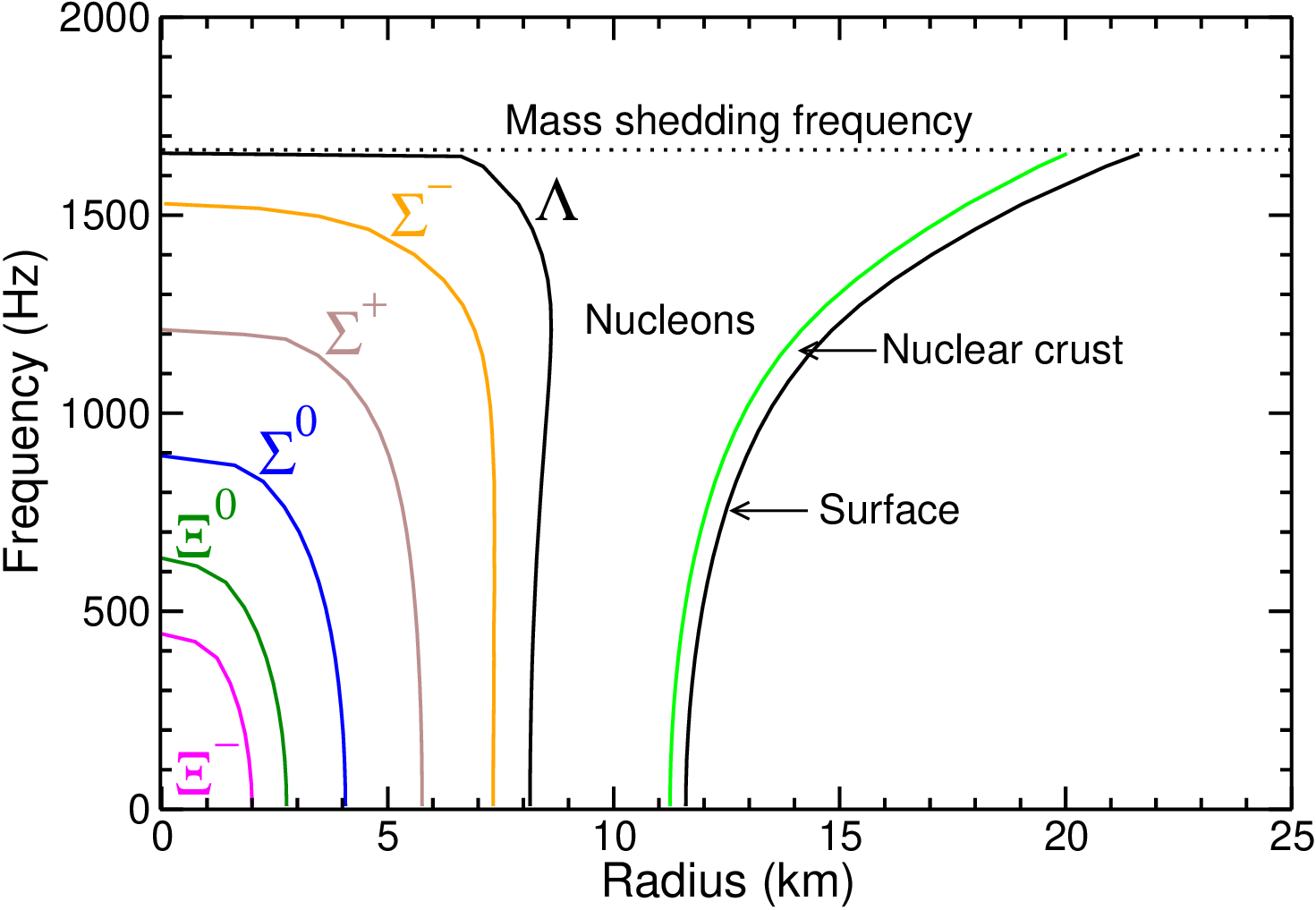,width=7.0cm}
{\caption[]{Hyperon composition of a rotating neutron star in
equatorial direction.}
\label{fig:bonn_eq}}}
\ \hskip 0.1cm \
\parbox[t]{7.5cm}
{\psfig{file=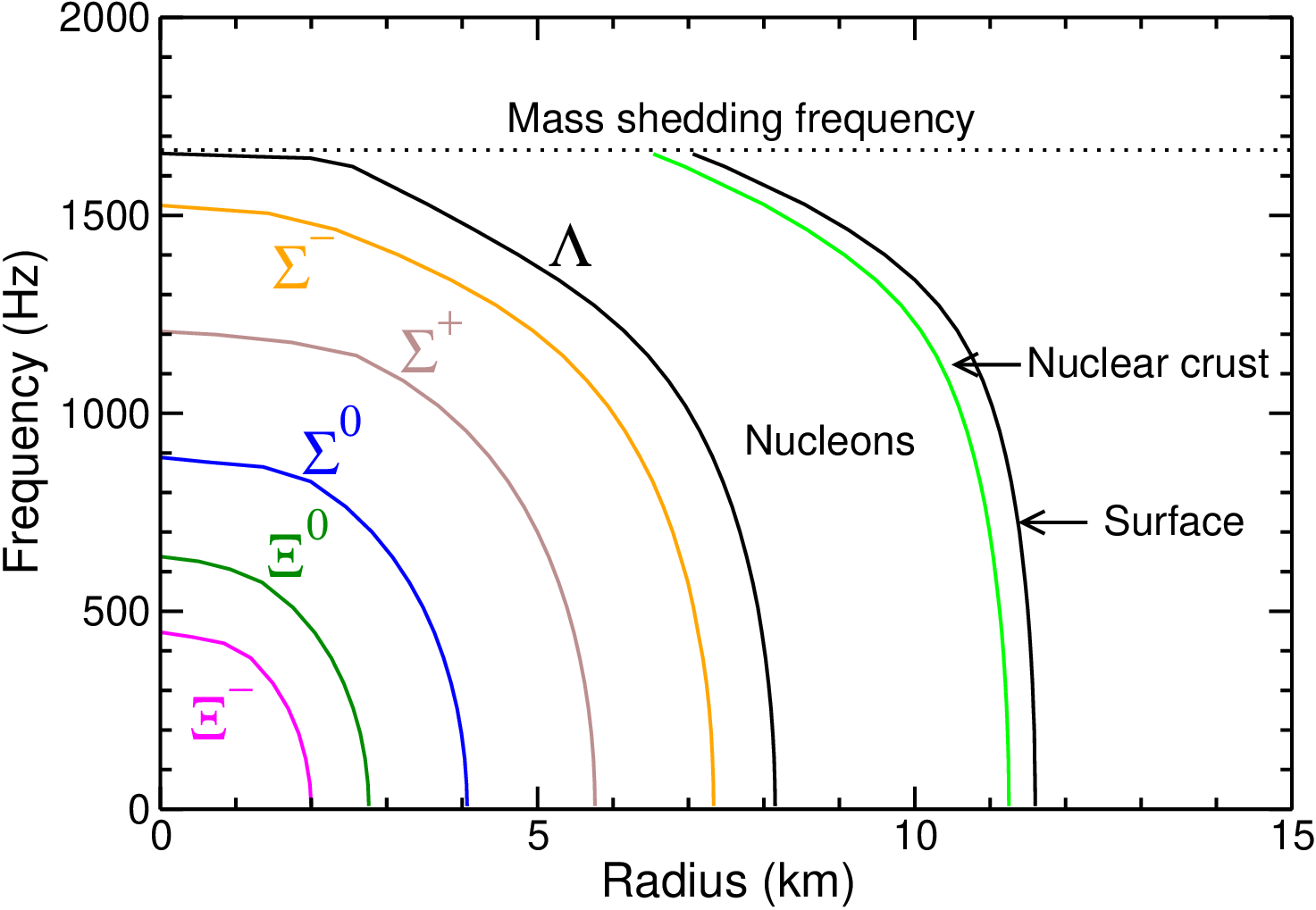,width=7.0cm} {\caption[]{
Same as Fig.\ \ref{fig:bonn_eq}, but in polar direction.}
\label{fig:bonn_po}}}
\end{center}
\end{figure}

\begin{figure}[tb]
\begin{center}
\parbox[t]{7.5cm} {\psfig{file=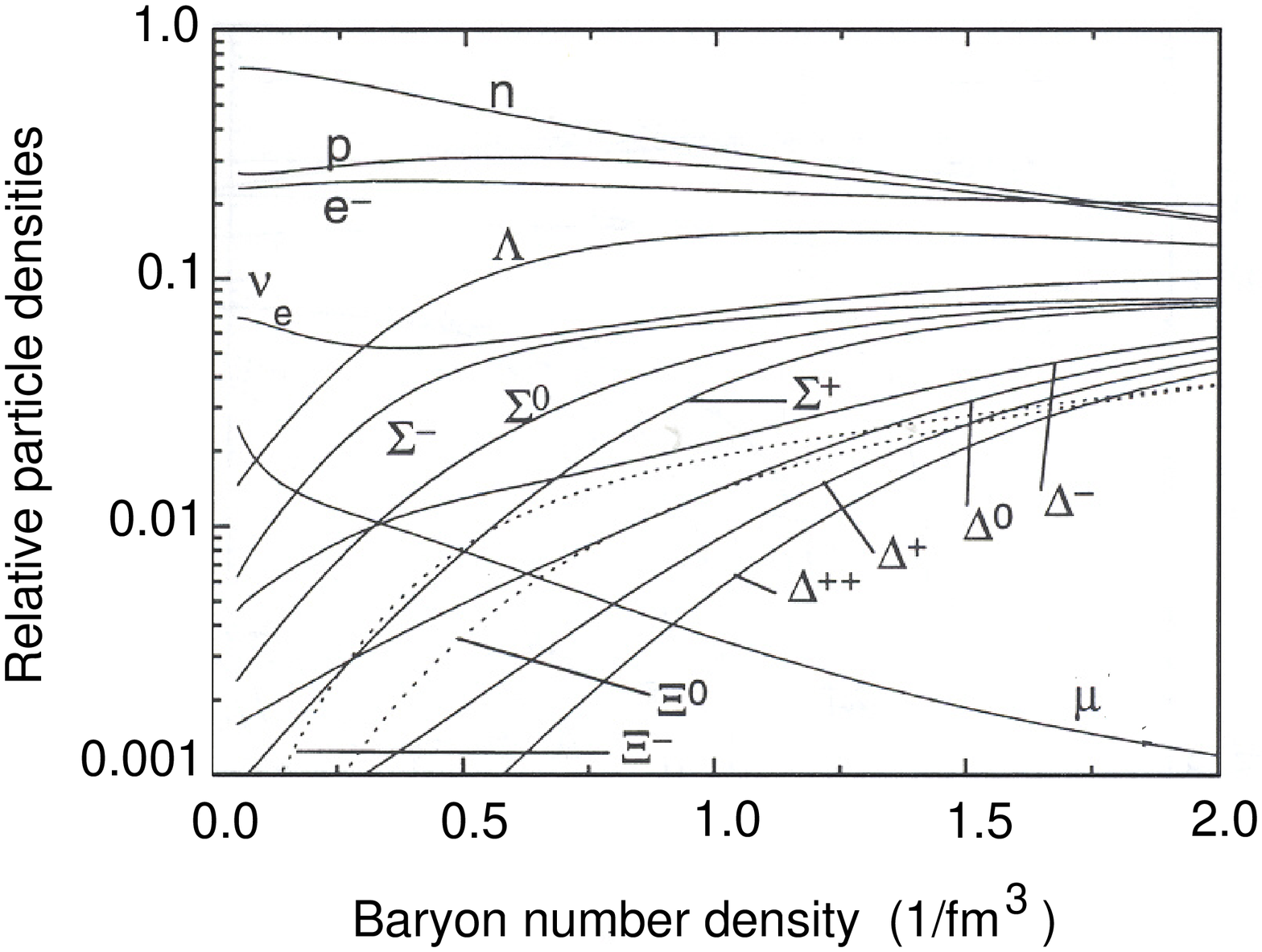,width=7.0cm}
{\caption[]{Composition of hot ($T=40$~MeV) proto-neutron star matter
for $Y_L = 0.3$ \cite{weber06:iship}.}
\label{fig:hotpnsm}}}
\ \hskip 0.1cm \
\parbox[t]{7.5cm}
{\psfig{file=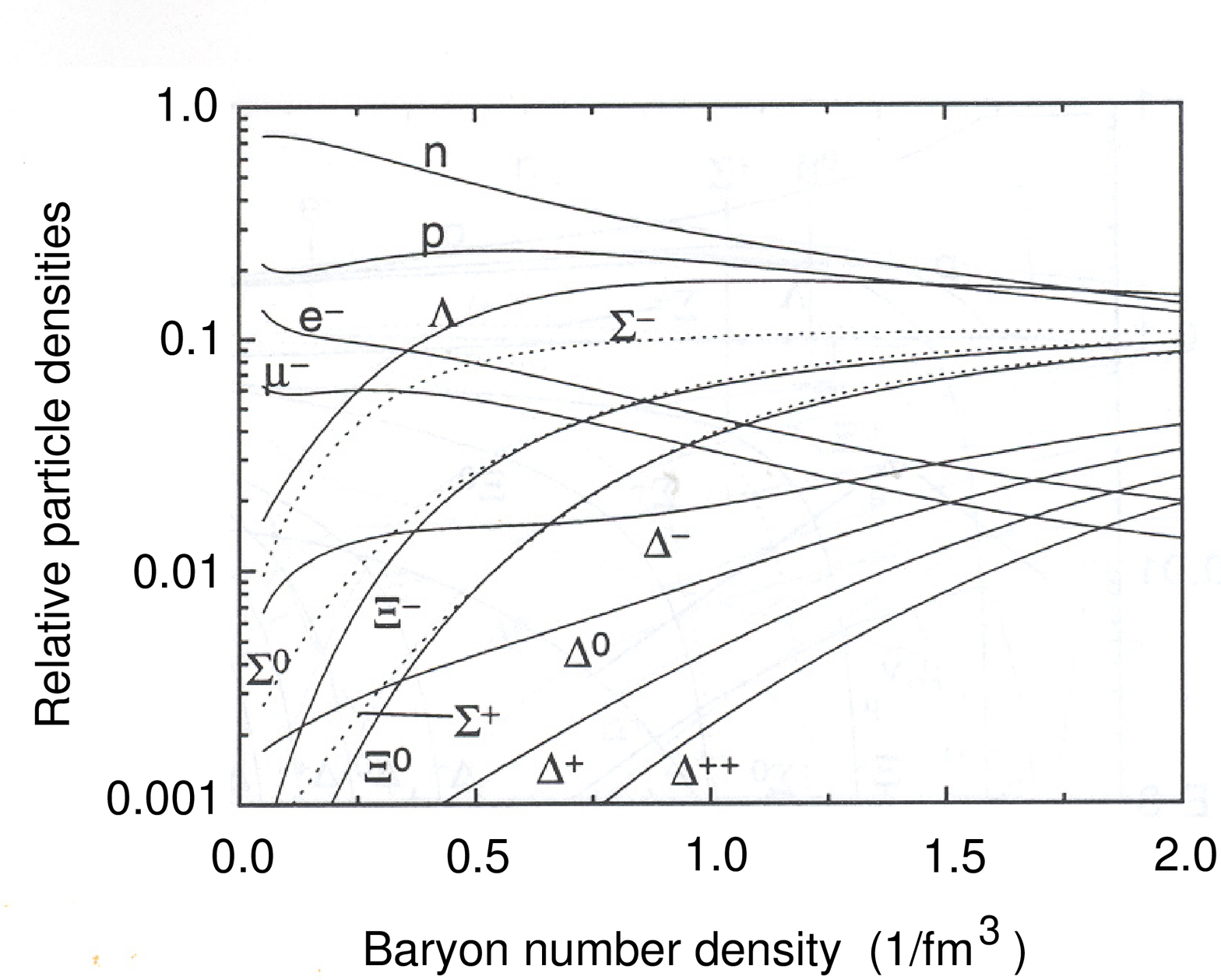,width=7.0cm} {\caption[]{
Same as Fig.\ \ref{fig:hotpnsm}, but for standard ($Y_L = 0$)
neutron star matter \cite{weber06:iship}.}
\label{fig:hotnsm}}}
\end{center}
\end{figure}

At the densities in the interior of neutron stars, the neutron
chemical potential, $\mu^n$, is likely to exceed the masses, modified
by interactions, of $\Sigma,~ \Lambda$ and possibly $\Xi$ hyperons
\cite{weber99:book,glen85:b}. Hence, in addition to nucleons, neutron
star matter may be expected to contain significant populations of
strangeness carrying hyperons.  The thresholds of the lightest baryon
resonances ($\Delta^-, \Delta^0, \Delta^+, \Delta^{++}$) are not
reached in
\begin{figure}[tb]
\begin{center}
\parbox[t]{7.5cm}
{\psfig{figure=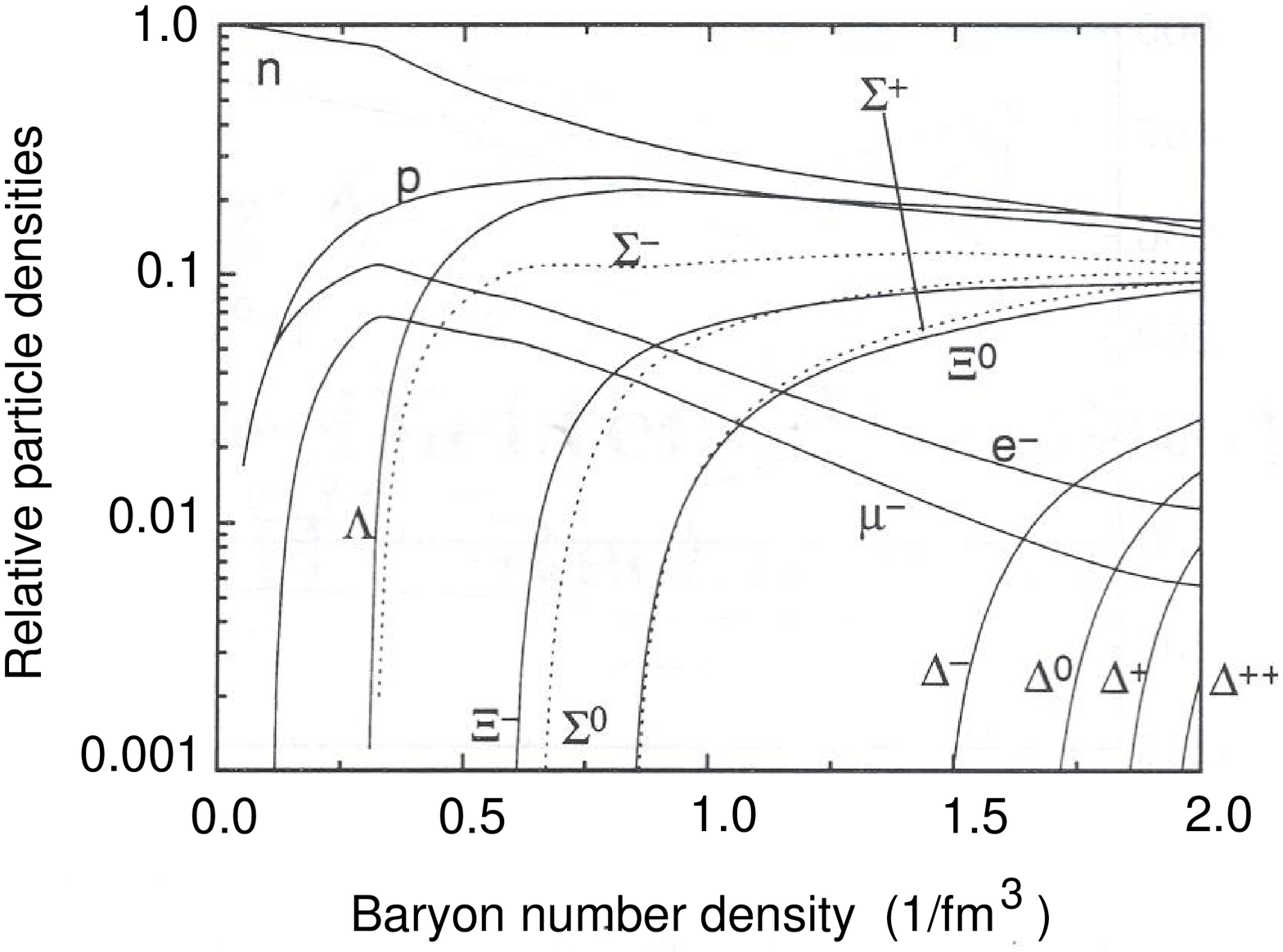,width=7.0cm,angle=0}
{\caption[]{Composition of standard neutron star matter, i.e.\
$T=0$~MeV and $Y_L = 0$ for the RMF approximation.}
\label{fig:coldnsm}}}
\ \hskip 0.1cm \
\parbox[t]{7.5cm} {\psfig{file=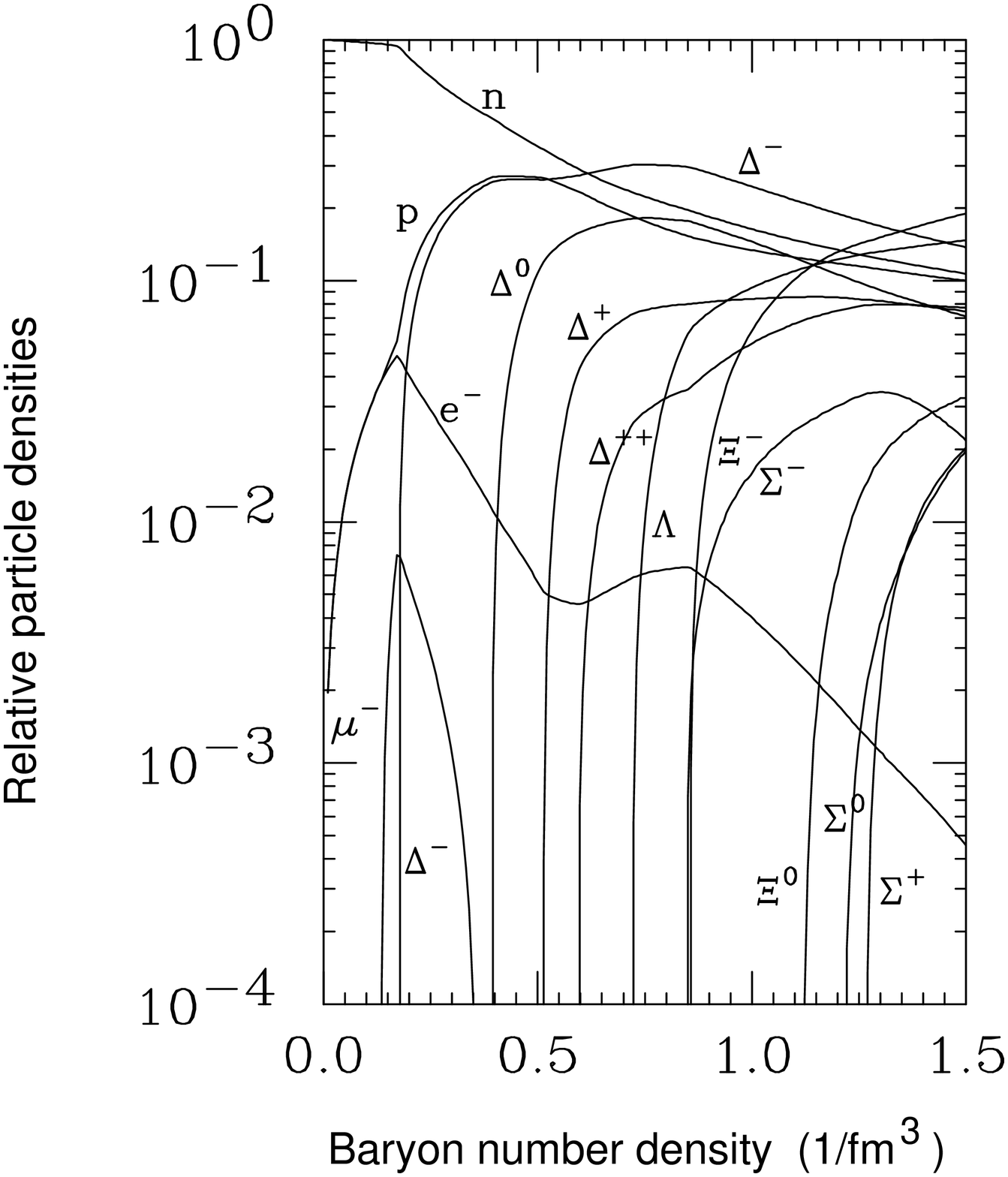,width=6.0cm,angle=0}
{\caption[]{Same as Fig.\ \ref{fig:coldnsm}, but computed for the RBHF
approximation \protect{\cite{huber98:a}}.}
\label{fig:9.Hub97a}}}
\end{center}
\end{figure}
relativistic mean-field (RMF) calculations. This is different for
many-body calculations performed for the relativistic
Brueckner-Hartree-Fock (RBHF) approximation where $\Delta$'s appear
rather abundantly \cite{huber98:a}, compare Figs.\ \ref{fig:coldnsm}
and \ref{fig:9.Hub97a}. Depending on the star's mass, the total
hyperon population can be very large \cite{glen85:b}, which is
illustrated graphically in Figs.\ \ref{fig:bonn_eq} and
\ref{fig:bonn_po} for rotating neutron stars whose EoS
is computed in the framework of the relativistic DD-RBHF formalism
\cite{hofmann01:a}. The stars shown in these figures have rotational
frequencies ranging from zero to the mass shedding frequency, $\nuk$, which is
the maximum frequency a star can have before mass loss at the equator
sets in. This frequency sets an absolute upper limit on stable rapid
rotation and follows from  a metric of the form
\cite{weber99:book,friedman86:a}
\begin{eqnarray}
  ds^2 = - \, e^{2\,\nu} \, d t^2 + e^{2\,\psi} \,
  \left( d\phi - \omega \, dt \right)^2 + e^{2\,\mu} \,
  d\theta^2 + e^{2\,\lambda} \, dr^2 \, ,
\label{eq:f220.exact} 
\end{eqnarray} 
where $\nu$, $\psi$, $\mu$ and $\lambda$ denote the metric functions
which depend on the radial coordinate $r$, polar angle $\theta$ and,
implicitly, on the star's angular velocity $\Omega$ as well as on the
angular velocity $\omega$ at which the local inertial frames are
dragged along in the direction of the star's rotation. In Newtonian
mechanics, mass shedding is determined by the equality between
centrifuge and gravity, and is readily obtained as $2 \pi \nuk =
\sqrt{M/R^3}$. Its general relativistic counterpart, computed here, is
given by \cite{weber99:book,friedman86:a}
\begin{eqnarray}
  2 \pi \nuk = \omega +\frac{\omega^\prime}{2\psi^\prime} +e^{\nu
  -\psi} \left( \frac{\nu^\prime}{\psi^\prime} +
  \left(\frac{\omega^\prime}{2 \psi^\prime}e^{\psi-\nu}\right)^2
  \right)^{1/2} \, .
\label{eq:okgr}  
\end{eqnarray} The primes denote derivatives with respect to the
Schwarzschild radial coordinate.  Finally, we take a brief look at the
composition of proto-neutron star matter. The composition of such
matter is determined by the requirements of charge neutrality and
equilibrium under the weak processes, $B_1 \rightarrow B_2 + l +
\bar\nu_l$ and $B_2 + l \rightarrow B_1 + \nu_l$, where $B_1$ and
$B_2$ are baryons, and $l$ is a lepton, either an electron or a
muon. For standard neutron star matter, where the neutrinos have left
the system, these two requirements imply that $Q = \sum_{i} q_i
n_{B_i} + \sum_{l=e, \mu} q_l n_l = 0$ (electric charge neutrality)
and $\mu^{B_i} = b_i \mu^n - q_i \mu^l$ (chemical equilibrium), where
$q_{i/l}$ denotes the electric charge density of a given particle, and
$n_{B_i}$ ($n_l$) is the baryon (lepton) number density.  The
subscript $i$ runs over all the baryons considered. The symbol
$\mu^{B_i}$ refers to the chemical potential of baryon $i$, $b_i$ is
the particle's baryon number, and $q_i$ is its charge. The chemical
potential of the neutron is denoted by $\mu^n$.  When the neutrinos
are trapped, as it is the case for proto-neutron star matter, the
chemical equilibrium condition is altered to $\mu^{B_i} = b_i \mu^n -
q_i (\mu^l - \mu^{\nu_l})$ and $\mu^e - \mu^{\nu_e} = \mu^\mu -
\mu^{\nu_\mu}$, where $\mu^{\nu_l}$ is the chemical potential of the
neutrino $\nu_l$. In proto-neutron star matter, the electron lepton
number $Y_L = (n_e+n_{\nu_e})/n_B$ is initially fixed at a value of
around $Y_{L_e} = Y_e + Y_{\nu_e} \simeq 0.3 - 0.4$ as suggested by
gravitational collapse calculations of massive stars. Also, because no
muons are present when neutrinos are trapped, the constraint
$Y_{L_\mu} = Y_\mu + Y_{\nu_mu} =0$ can be imposed. Figures
\ref{fig:hotpnsm} and \ref{fig:hotnsm} show sample compositions of
proto-neutron star matter and standard neutron star matter (no
neutrinos) computed for the relativistic mean-field approximation. The
presence of the $\Delta$ particle in (proto) neutron star matter at
finite temperature is striking. As already mentioned at the beginning
of this section, this particle is generally absent in cold neutron
star matter treated in the relativistic mean-field approximation.

\goodbreak
\subsection{\it Meson condensation}\label{ssec:mcondens}

The condensation of negatively charged mesons in neutron star matter
is favored because such mesons would replace electrons with very high
Fermi momenta. Early estimates predicted the onset of a negatively
charged pion condensate at around $2 n_0$. However, these estimates
are very sensitive to the strength of the effective nucleon
particle-hole repulsion in the isospin $T=1$, spin $S=1$ channel,
described by the Landau Fermi-liquid parameter $g'$, which tends to
suppress the condensation mechanism. Measurements in nuclei tend to
indicate that the repulsion is too strong to permit condensation in
nuclear matter \cite{barshay73:a,brown88:a}. In the mid 1980s, it was
discovered that the in-medium properties of $K^- [u \bar s]$ mesons
may be such that this meson rather than the $\pi^-$ meson may condense
in neutron star matter \cite{kaplan86:a,brown87:a}.  The condensation
is initiated by the schematic reaction $e^- \rightarrow K^- + \nu_e$.
If this reaction becomes possible in neutron star matter, it is
energetically advantageous to replace the fermionic electrons with the
bosonic $K^-$ mesons. Whether or not this happens depends on the
behavior of the $K^-$ mass, $m^*_{K^-}$, in neutron star matter.
Experiments which shed light on the properties of the $K^-$ in nuclear
matter have been performed with the Kaon Spectrometer (KaoS) and the
FOPI detector at the heavy-ion synchrotron SIS at GSI
\cite{fuchs06:a}.  An analysis of the early $K^-$ kinetic energy
spectra extracted from Ni+Ni collisions showed that the attraction
from nuclear matter would bring the $K^-$ mass down to
$m^*_{K^-}\simeq 200~\mev$ at densities $\sim 3\, n_0$. For
neutron-rich matter, the relation $m^*_{K^-} / m_{K^-} \simeq 1 - 0.2
n / n_0$ was established \cite{brown97:a}, with $m_K = 495$~MeV the
$K^-$ vacuum mass.  Values of around $m^*_{K^-}\simeq 200~\mev$ may be
reached by the electron chemical potential, $\mu^e$, in neutron star
matter \cite{weber99:book,glen85:b} so that the threshold condition
for the onset of $K^-$ condensation, $\mu^e = m^*_K$, might be
fulfilled for sufficiently dense neutron stars, provided other
negatively charged particles ($\Sigma^-$, $\Delta^-$, $d$ and $s$
quarks) are not populated first and prevent the electron chemical
potential from increasing with density.  We also note that $K^-$
condensation allows the conversion reaction $n \rightarrow p +
K^-$. By this conversion the nucleons in the cores of neutron stars
can become half neutrons and half protons, which lowers the energy per
baryon of the matter. The relative isospin symmetric composition
achieved in this way resembles the one of atomic nuclei, which are
made up of roughly equal numbers of neutrons and protons.  Neutron
stars are therefore referred to, in this picture, as nucleon
stars. The maximum mass of such stars has been calculated to be around
$1.5\, \msun$ \cite{thorsson94:a}. Consequently, the collapsing core
of a supernova, e.g.\ 1987A, if heavier than this value, should go
into a black hole rather than forming a neutron star, as pointed out
by Brown et al.\ \cite{brown94:a}. This would imply the existence of a
large number of low-mass black holes in our galaxy \cite{brown94:a}.
Thielemann and Hashimoto \cite{thielemann90:a} deduced from the total
amount of ejected $^{56}{\rm Ni}$ in supernova 1987A a neutron star
mass range of $1.43 - 1.52~ \msun$. If the maximum neutron star mass
should indeed be in this mass range ($\sim 1.5~\msun$),  
the existence of heavy neutron stars with masses
around $2~\msun$ (Sect.\ \ref{sec:masses}) would be ruled out. 

\subsection{\it H-matter and exotic baryons}

A novel particle that could be of relevance for the composition of
neutron star matter is the H-dibaryon (H=$([ud][ds][su])$), a doubly
strange six-quark composite with spin and isospin zero, and baryon
number two \cite{jaffe77:a}. Since its first prediction in 1977, the
H-dibaryon has been the subject of many theoretical and experimental
studies as a possible candidate for a strongly bound exotic state. In
neutron star matter, which may contain a significant fraction of
$\Lambda$ hyperons, the $\Lambda$'s could combine to form H-dibaryons,
which could give way to the formation of H-dibaryon matter at
densities somewhere above $\sim 4\, n_0$ \cite{glen98:a}.  If formed
in neutron stars, however, H-matter appears unstable against
compression which could trigger the conversion of neutron stars into
hypothetical strange stars \cite{faessler97:a,faessler97:b}.  Another
particle, referred to as an exotic baryon, of potential relevance for
neutron stars, could be the pentaquark, $\Theta^+ ([ud]^2 \bar s)$,
with a predicted mass of 1540~MeV. The pentaquark, which carries
baryon number one, is a hypothetical subatomic particle consisting of
a group of four quarks and one anti-quark (compared to three quarks in
normal baryons and two in mesons), bound by the strong color-spin
correlation force (attraction between quarks in the color $\bar {\bf
3}_c$ channel) that drives color superconductivity
\cite{jaffe03:a}. The pentaquark decays according to $\Theta^+(1540)
\rightarrow K^+ [\bar s u] + n[udd]$ and thus has the same quantum
numbers as the $K^+ n$.

\subsection{\it Quark deconfinement}\label{ssec:deconf}

It has been suggested already several decades ago
\cite{ivanenko65:a,itoh70:a,fritzsch73:a,baym76:a,keister76:a,%
chap77:a,fech78:a,chap77:b} that the nucleons in the cores of neutron
stars may melt under the enormous pressures that exist in the cores,
creating a new state of matter know as quark matter. From simple
geometrical considerations it follows that quark confinement could
occur at densities somewhere between around $2-10\, n_0$. Depending on
rotational frequency and neutron star mass, densities greater than two
to three times $n_0$ are easily reached in the cores of neutron stars
so that the neutrons and protons in the cores of neutron stars may
indeed be broken up into their quarks constituents
\cite{glen97:book,weber99:book,weber05:a,glen91:pt}. More than that,
since the mass of the strange quark is only around 150~MeV,
high-energetic up and down quarks will readily transform to strange
quarks at about the same density at which up and down quark
deconfinement sets in. Thus, if quark matter exists in the cores of
\begin{figure}[tb]
\begin{center}
\epsfig{figure=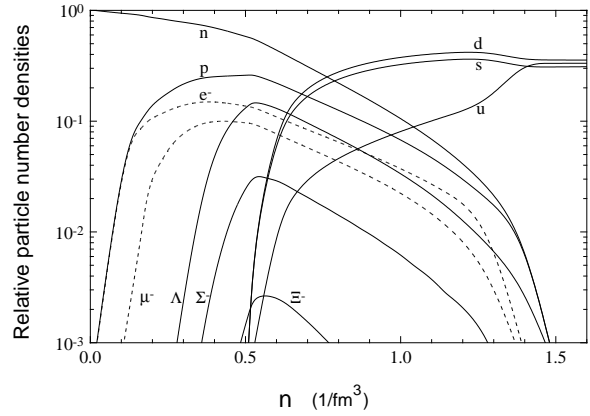,width=8.0cm,angle=0}
\begin{minipage}[t]{16.5 cm}
\caption{Sample composition of chemically equilibrated quark-hadron
(hybrid star) matter as a function of baryon number
density \cite{weber99:book,weber05:a}.}
\label{fig:3.13.1}
\end{minipage}
\end{center}
\end{figure}
neutron stars, it should be made of the three lightest quark flavors.
The remaining three quark flavors (charm, top, bottom) are way to
massive to be created in neutron stars. For instance, the creation of
charm quark requires a density greater than $10^{17}\, \gcmt$, which
around 100 times greater than the density reached in neutron stars.  A
stability analysis of stars with a charm quark population reveals that
such objects are unstable against radial oscillations and, thus, can
not exist stably in the Universe \cite{weber99:book,weber05:a}. The
same is true for ultra-compact stars with unconfined populations of
top and bottom quarks, since the pulsation eigen-equations are of
Sturm-Liouville type.

The phase transition from confined hadronic (H) matter to deconfined
quark (Q) matter is characterized by the conservation of baryon charge
and electric charge. The Gibbs condition for phase equilibrium then is
that the two associated chemical potentials, $\mu^n$ and $\mu^e$, and
the pressure in the two phases be equal \cite{glen91:pt}, $ P_{\rm
H}(\mu^n,\mu^e, \{ \psi_\chi \}, T) = P_{\rm Q}(\mu^n,\mu^e,T)$, where
$P_{\rm H}$ denotes the pressure of hadronic matter computed for a
given hadronic Lagrangian ${\cal L}_{\rm M}(\{\psi_\chi\})$, with
$\{\psi_\chi\}$ the field variables and Fermi momenta that characterize a
solution to the field equations of confined hadronic matter,
\begin{eqnarray}
  ( i \gamma^\mu\partial_\mu - m_\chi ) \psi_\chi(x) &=&
  \sum_{M=\sigma,\omega,\pi, ...} \hat\Gamma_{M \chi} M(x) \,
  \psi_\chi(x) \, , \\ ( \partial^\mu\partial_\mu + m^2_\sigma)
  \sigma(x) &=& \sum_{\chi = p, n, \Sigma, ...} \hat\Gamma_{\sigma
  \chi}\, \bar\psi_\chi(x) \psi_\chi(x) \, , \qquad {\rm etc.}
\end{eqnarray}
The pressure of quark matter, $P_{\rm Q}$, is obtainable from the bag
model. The quark chemical potentials $\mu^u, ~\mu^d, ~\mu^s$ are
related to the baryon and charge chemical potentials as $\mu^u =
(\mu^n - 2 \mu^e) / 3$ and $\mu^d = \mu^s = (\mu^n + \mu^e) /3$. The
Gibbs condition is to be supplemented with the global relations for
conservation of baryon charge and electric charge within an unknown
volume $V$ containing $A$ baryons. The first one is given by $n \equiv
A/ V = (1-\chi) n_{\rm H}(\mu^n,\mu^e,T) + \chi n_{\rm
Q}(\mu^n,\mu^e,T)$, where $\chi \equiv V_{\rm Q}/V$ denotes the volume
proportion of quark matter, $V_{\rm Q}$, in the unknown volume $V$,
and $n_{\rm H}$ and $n_{\rm Q} $ are the baryon number densities of
hadronic matter and quark matter.  Global neutrality of electric
charge within the volume $V$ can be written as $0 = Q/V = (1-\chi) 
q_{\rm H}(\mu^n,\mu^e,T) + \chi q_{\rm Q}(\mu^n,\mu^e,T) + q_{\rm
L}$, with $q_i$ the electric charge densities of hadrons, quarks, and
leptons.  For a given temperature, $T$, these equations serve to
determine the two independent chemical potentials and the volume $V$
for a specified
\begin{figure}[tb]
\begin{center}
\includegraphics*[width=0.9\textwidth,angle=0,clip]{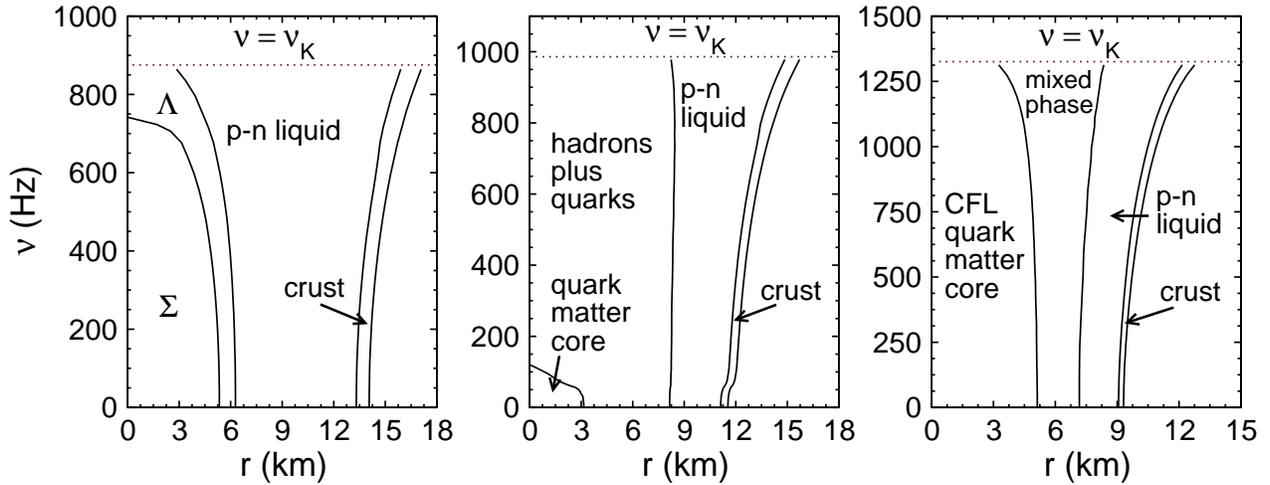}
\begin{minipage}[t]{16.5 cm}
\caption{Dependence of neutron star composition on spin frequency,
$\nu$, for three sample compositions (left: hyperon composition,
middle: quark-hybrid composition, right: quark-hybrid composition with
quark matter in the color-flavor locked phase). The non-rotating
stellar mass in each case is $1.4\, \msun$. $\nuk$ denotes the Kepler
(mass-shedding) frequency.}
\label{fig:profiles}
\end{minipage}
\end{center}
\end{figure}
volume fraction $\chi$ of the quark phase in equilibrium with the
hadronic phase.  After completion $V_{\rm Q}$ is obtained as $V_{\rm
Q}=\chi V$. The chemical potentials depend on the proportion $\chi$ of
the phases in equilibrium, and hence so do all properties that
depend on them, i.e.\ the energy densities, baryon and charge
densities of each phase, and the common pressure. For the mixed phase,
the volume proportion of quark matter varies from $0 \leq \chi \leq 1$
and the energy density is the linear combination of the two phases
$\epsilon = (1-\chi) \epsilon_{\rm H}(\mu^n,\mu^e, \{\psi_\chi\}, T) +
\chi \epsilon_{\rm Q}(\mu^n,\mu^e,T)$. Model neutron star compositions
computed within the framework described just above are shown in Figs.\
\ref{fig:3.13.1} and \ref{fig:profiles}.  Possible astrophysical
signals associated with quark deconfinement, the most striking of
which being ``backbending'' of isolated pulsars, are discussed in
\cite{glen97:book,weber99:book,weber05:a,chubarian00:a}.

\goodbreak
\subsection{\it Color-superconductivity of quark matter}\label{sec:color}

There has been much recent progress in our understanding of quark
matter, culminating in the discovery that if quark matter exists it
ought to be in a color superconducting state
\cite{alford98:a,rajagopal01:a,alford01:a,rapp98+99:a}. This is made
possible by the strong interaction among the quarks which is very
attractive in some channels (antisymmetric antitriplet
channel). Pairs of quarks are thus expected to form Cooper pairs very
readily. Since pairs of quarks cannot be color-neutral, the resulting
condensate will break the local color symmetry and form what is called
a color superconductor.  The phase diagram of such matter is expected
to be very complex \cite{rajagopal01:a,alford01:a}. The complexity is
caused by the fact that quarks come in three different colors,
different flavors, and different masses.  Moreover, bulk matter is
neutral with respect to both electric and color charge, and is in
chemical equilibrium under the weak interaction processes that turn
one quark flavor into another. To illustrate the condensation pattern
briefly, we note the following pairing ansatz for the quark condensate
$ \langle \psi^\alpha_{f_a} C\gamma_5 \psi^\beta_{f_b} \rangle \sim
\Delta_1 \epsilon^{\alpha\beta 1}\epsilon_{{f_a}{f_b}1} + \Delta_2
\epsilon^{\alpha\beta 2}\epsilon_{{f_a}{f_b}2} + \Delta_3
\epsilon^{\alpha\beta 3}\epsilon_{{f_a}{f_b}3}$, where
$\psi^\alpha_{f_a}$ is a quark of color $\alpha=(r,g,b)$ and flavor
${f_a}=(u,d,s)$. The condensate is a Lorentz scalar, antisymmetric in
Dirac indices, antisymmetric in color, and thus antisymmetric in
flavor. The gap parameters $\Delta_1$, $\Delta_2$ and $\Delta_3$
describe $d$-$s$, $u$-$s$ and $u$-$d$ quark Cooper pairs,
respectively. The following pairing schemes have emerged: At
asymptotic densities ($m_s \rightarrow 0$ or $\mu \rightarrow \infty$)
the ground state of QCD with a vanishing strange quark mass is the
color-flavor locked (CFL) phase (color-flavor locked quark pairing),
in which all three quark flavors participate symmetrically.  The gaps
associated with this phase are $\Delta_3 \simeq \Delta_2 = \Delta_1 =
\Delta$, and the quark condensates of the CFL phase are approximately
of the form $ \langle \psi^{\alpha}_{f_a} C\gamma_5 \psi^{\beta}_{f_b}
\rangle \sim \Delta \, \epsilon^{\alpha \beta X} \epsilon_{{f_a} {f_b}
X}$, with color and flavor indices all running from 1 to 3. Since
$\epsilon^{\alpha\beta X} \epsilon_{{f_a} {f_b} X} =
\delta^\alpha_{f_a}\delta^\beta_{f_b} -
\delta^\alpha_{f_b}\delta^\beta_{f_a}$ one sees that the condensate
involves Kronecker delta functions that link color and flavor
indices. Hence the notion color-flavor locking. The CFL phase has been
shown to be electrically neutral without any need for electrons for a
significant range of chemical potentials and strange quark masses
\cite{rajagopal01:b}. If the strange quark mass is heavy enough to be
ignored, then up and down quarks may pair in the two-flavor
superconducting (2SC) phase.  Other possible condensation patterns are
CFL-$K^0$, CFL-$K^+$ and CFL-$\pi^{0,-}$, gCFL (gapless CFL phase),
1SC (single-flavor-pairing), CSL (color-spin locked phase), and the
LOFF (crystalline pairing) phase, depending on $m_s$, $\mu$, and
electric charge density (for references, see \cite{weber05:a}).
Calculations performed for massless up and down quarks and a very
heavy strange quark mass ($m_s \rightarrow \infty$) agree that the
quarks prefer to pair in the two-flavor superconducting (2SC) phase
where $ \Delta_3 > 0\, , \quad {\rm and} \quad \Delta_2 = \Delta_1 =
0$.  In this case the pairing ansatz  reduces
to $ \langle \psi^{\alpha}_{f_a} C \gamma_5 \psi^{\beta}_{f_b} \rangle
\propto \Delta \, \epsilon_{ab} \epsilon^{\alpha \beta 3}$.  Here the
resulting condensate picks a color direction (3 or blue in the example
above), and creates a gap $\Delta$ at the Fermi surfaces of quarks
with the other two out of three colors (red and green). The gapless
CFL phase (gCFL) may prevail over the CFL and 2SC phases at
intermediate values of $m^2_s/\mu$ with gaps given obeying the
relation $\Delta_3 > \Delta_2 > \Delta_1 > 0.$ For chemical potentials
that are of astrophysical interest, $\mu < 1000$~MeV, the gap is
between 50 and 100~MeV. The order of magnitude of this result agrees
with calculations based on phenomenological effective interactions
\cite{rapp98+99:a,alford99:b} as well as with perturbative
calculations for $\mu > 10$~GeV \cite{son99:a}. We also note that
superconductivity modifies the EoS at the order of $(\Delta / \mu)^2$
\cite{alford03:b,alford04:a}, which is even for such large gaps only a
few percent of the bulk energy. Such small effects may be safely
neglected in present determinations of models for the EoS of
quark-hybrid stars. There has been much recent work on how color
superconductivity in neutron stars could affect their properties
\cite{rajagopal01:a,alford01:a,alford00:a,rajagopal00:a,alford00:b,%
blaschke99:a}.  These studies reveal that possible signatures include
the cooling by neutrino emission, the pattern of the arrival times of
supernova neutrinos, the evolution of neutron star magnetic fields,
rotational stellar instabilities, and glitches in rotation
frequencies.

\goodbreak
\subsection{Absolutely stable strange quark matter}\label{sec:sqm}

It is most intriguing that for strange quark matter made of more than
a few hundred up, down, and strange quarks, the energy of strange
quark matter may be well below the energy of nuclear matter, $E/A=
930$~MeV, which gives rise to new and novel classes of strange matter
objects, ranging from strangelets at the low baryon-number end to
strange stars at the high baryon number end
\cite{weber99:book,weber05:a,alcock86:a,madsen98:b,alford06:a}.  The
presence of electrons in strange quark matter may lead to the
formation of an electric dipole layer on the surface of strange
matter, with huge electric fields on the order of $10^{19}$~V/cm. This
peculiar feature enables strange quark stars to be enveloped in 
nuclear crusts made of ordinary atomic nuclei
\cite{alcock86:a,stejner05:a,kettner94:b}. Sequences of compact strange
stars with and without (bare) nuclear crusts are shown in Fig.\
\ref{fig:MvsR}. Since the nuclear crust is gravitationally bound to
the quark matter, the mass-radius relationship of strange stars with
crusts resembles the one of neutron stars and even that of white
dwarfs \cite{glen94:a,mathews06:a}. In contrast to neutron stars,
however, strange stars obey $M \propto R^3$ because they are
self-bound and the mass density of quark matter is almost constant
inside strange stars.
The electrons surrounding strange quark matter are held to quark
matter electrostatically. Since neither component, electrons nor quark
matter, is held in place gravitationally, the Eddington limit to the
luminosity that a static surface may emit does not apply, and thus the
object may have photon luminosities much greater than $10^{38}~\ergs$.
It was shown by Usov \cite{usov98:a} that this value may be exceeded by
many orders of magnitude by the luminosity of $e^+ e^-$ pairs produced
by the Coulomb barrier at the surface of a hot strange star. For a
surface temperature of $\sim 10^{11}$~K, the luminosity in the
outflowing pair plasma was calculated to be as high as $\sim 3 \times
10^{51}~\ergs$.  Such an effect may be a good observational signature
of bare strange stars \cite{usov98:a,usov01:c,usov01:b,cheng03:a}. If
the strange star is enveloped by a nuclear crust however, which is
gravitationally bound to the strange star, the surface made up of
ordinary atomic matter would be subject to the Eddington limit. Hence
the photon emissivity of such a strange star would be the same as for
an ordinary neutron star.  If quark matter at the stellar surface is
in the CFL phase the process of $e^+ e^-$ pair creation at the stellar
quark matter surface may be turned off, since cold CFL quark matter is
electrically neutral so that no electrons are required and none are
admitted inside CFL quark matter \cite{rajagopal01:b}. This may be
different for the early stages of a hot CFL quark star \cite{vogt03:a}.

\section{Latent Heat of Phase Transitions}\label{sec:lheat}

A neutron star is born with enormous reserves of rotational and
thermal energy, which it looses over millions of years through
processes such as magnetic dipole radiation from the magnetosphere,
thermal radiation and electron winds from the surface and neutrinos
from the core. In doing so the star cools down and changes structure
as the radius diminishes with the rotation frequency--by as much as a
few kilometers if it started out near its limiting Kepler
frequency--shifting the interior boundaries of any phase transitions
there in the process. Any latent heat evolved or absorbed by particles
crossing these boundaries will contribute to the thermal evolution af
the star, and in this section we discuss the formalism for describing
these effects.
The thermal evolution of a spherical star in General Relativity is
given by a local energy balance equation \cite{thorne77:a,yakovlev04:a}:
\begin{equation}
\frac{e^{-\lambda-2\phi}}{4\pi r^2}\frac{\partial (L_r
  e^{2\phi})}{\partial r} = \epsilon_\mathrm{nuc}-\epsilon_\nu - \rho
  e^{-\phi}\frac{\partial \Pi}{\partial
  t}+e^{-\phi}\frac{P}{\rho}\frac{\partial \rho}{\partial t}\;,\qquad
  L_r=4\pi\kappa r^2
  e^{-\lambda-\phi}\frac{\partial(Te^\phi)}{\partial r}\label{balance}
  \ ,
\end{equation}
where $L_r$ is the local luminosity given by the conduction of
non-neutrino energy through a sphere of radius $r$, $\epsilon_{nuc}$
is the rate, per volume, at which nuclear reactions create non-nuclear
energy, $\epsilon_\nu$ is the neutrino emissivity per volume, $\rho$
is the rest mass density, $P$ is the pressure and $\Pi$ is the
specific internal energy per unit rest mass, which in the absence of
any phase transitions may be written
\begin{equation}
\Pi=\frac{1}{\rho}\int_0^T c_\mathrm{V}(\rho,T') \,\mathrm{d}T'\;,
\end{equation}
with $c_\mathrm{V}$ the heat capacity per unit volume and $T$ the
temperature. Since $P=-(\partial E/\partial V)_T $ we then have
$\partial \Pi/\partial \rho=P/\rho^2$ so in this case
\begin{equation}
  \frac{\partial \Pi}{\partial t}=\frac{\partial\Pi}{\partial
    T}\frac{\partial T}{\partial t} +\frac{\partial\Pi}{\partial
    \rho}\frac{\partial \rho}{\partial
    t}=\frac{c_\mathrm{V}}{\rho}\frac{\partial T}{\partial t}
    +\frac{P}{\rho^2}\frac{\partial \rho}{\partial t} \ ,
\end{equation}
which gives the usual heat balance equation for neutron stars
\begin{equation}
\frac{e^{-\lambda-2\phi}}{4\pi r^2}\frac{\partial (L_r
  e^{2\phi})}{\partial r} = \epsilon_\mathrm{nuc}-\epsilon_\nu -
  e^{-\phi}c_\mathrm{V}\frac{\partial T}{\partial t}\; .
\end{equation}
However in the presence of any phase transitions the heat capacity may
be discontinuous and this will give rise to extra terms as particles
cross the phase boundaries in either temperature for superfluid
transitions or in density if there is a transition from hadronic
matter to quark matter.  To see this for a sharp phase transition with
no mixed phase, we may write $\Pi$ as
\begin{equation}
\Pi=\rho^{-1}\left[\Theta(1-\lambda/\lambda_{c,1})\int_0^Tc_\mathrm{V,1}
  \mathrm{d}T'+\Theta(\lambda/\lambda_{c,2}-1)\int_0^T
  c_{\mathrm{V,2}}\mathrm{d}T'\right]\;,
\end{equation}  
where $\Theta$ is the Heaviside step function, $\lambda$ can be either
$T$, $\rho$ (which may be discontinous at the phase transition) or
some other convenient quantity depending on the nature of the phase
transition, $\lambda_c$ is the critical $\lambda$ for the phase
transition and numerical subscripts identifies a variable with either
of the two phases. Differentiating with respect to time we then get
the same terms as above and additional terms from the Heaviside
functions:
\begin{equation}
  \frac{\partial \Pi}{\partial
    t}=\frac{c_\mathrm{V}}{\rho}\frac{\partial T}{\partial
    t}+\frac{P}{\rho^2}\frac{\partial \rho}{\partial t} + \frac
    {1}{\rho}\left[\frac{\delta(\lambda/\lambda_{c,2}-1)}{\lambda_{c,2}}
    \frac{\partial \lambda}{\partial t}
    \int_0^Tc_{\mathrm{V,2}}\mathrm{d}T'-\frac{\delta(\lambda/
    \lambda_{c,1}-1)}{\lambda_{c,1}}\frac{\partial \lambda}{\partial
    t}\int_0^Tc_{\mathrm{V,1}}\mathrm{d}T'\right]\;.
\end{equation}
Simplifying this we note that $c_\mathrm{V}=T(\partial s / \partial
T)_\mathrm{V}$ where $s$ is the entropy density, integrate by parts
and use the phase equilibrium conditions that $P_1=P_2, T_1=T_2,
\mu_1=\mu_2$ to get
\begin{equation}
  \frac{\partial \Pi}{\partial
    t}=\frac{c_\mathrm{V}}{\rho}\frac{\partial T}{\partial
    t}+\frac{P}{\rho^2}\frac{\partial \rho}{\partial t} + \frac
    {1}{\rho}\frac{\partial \lambda}{\partial
    t}T\left[\frac{\delta(\lambda/\lambda_{c,2}-1)}{\lambda_{c,2}}s_2
    -
    \frac{\delta(\lambda/\lambda_{c,1}-1)}{\lambda_{c,1}}s_1\right]\;. 
\label{sharp}
\end{equation}
This is of course the well known result that each particle crossing a
phase boundary contributes to the surrounding medium an energy
\cite{landau80:a} $q=-T[\sigma_2-\sigma_1]=-T\Delta\sigma$, where
$\sigma$ is the entropy per particle. If $\lambda$ is a continuous
variable such as $T$ or $P$ the distinction between $\lambda_{c,1}$
and $\lambda_{c,2}$ is pointless and the delta functions may be taken
outside, and a second order transition with continuous entropy density
at the phase transition would then cause the last term in Eq.\
(\ref{sharp}) to vanish.

A mixed phase may be treated similarly by replacing the Heaviside
functions with a volume fraction $\chi$, such that $1-\chi$ is the
fraction of a given volume occupied by phase 1 and $\chi$ is that
occupied by phase 2 so $\Pi=(1-\chi)\Pi_1+\chi\Pi_2$ and hence
\begin{equation}
  \frac{\partial \Pi}{\partial t}=\frac{c_\mathrm{V}}{\rho}\frac{\partial
    T}{\partial t}+\frac{P}{\rho^2}\frac{\partial \rho}{\partial t} +
  \frac
  {1}{\rho}\frac{\partial \chi}{\partial\lambda}\frac{\partial
    \lambda}{\partial
    t}T\left[s_2-s_1\right]\;. \label{mixed}
\end{equation}
The transition to a superfluid state is a second order transition
taking place through a mixed phase with $\Delta\sigma\propto
-(1-T/T_c)$ and $\lim_{T\rightarrow 0}\chi=
1-(2\pi\Delta_0/T)^{1/2}e^{-\Delta_0/T}$ for simple Fermi gasses
\cite{landau80:a}. The latent heat released in this process is
customarily treated as an increase in the heat capacity letting
$c_\mathrm{V}\rightarrow c_\mathrm{V}R(T/T_c)$ with the effect of
slightly slowing the cooling process -- at temperatures around
$0.2T_c$ the reduction factor, $R(T/T_c)$, goes to zero and has the
opposite effect of significantly accelerating the cooling. This has
been extensively treated in the literature and we refer to
\cite{yakovlev99:a} for a recent review.

The transition from hadronic to quark matter is a first order
transition which will take place through a mixed phase
\cite{glen91:pt} if finite size effects and charge screening
do not prohibit such a phase \cite{heiselberg92:a,endo06:a} 
(although see the comments in \cite{christiansen97:a}). If the
transition is sharp it happens at a specific critical pressure, where
the density will be discontinuous and heat will be released there at a
total rate of $-e^{-\phi}T\Delta\sigma
(\mathrm{d}N_\mathrm{q}/\mathrm{d}t)$, with
$\mathrm{d}N_\mathrm{q}/\mathrm{d}t$ the number of particles making
the transition per unit time. If the transition is through a mixed
phase the pressure and {\em average} entropy density, $s= (1-\chi)s_1
+\chi s_2$, vary continously in the region where the phase transition
takes place, but the individual entropy densities will {\em not} be
equal and particles making the transition will release heat at a local
rate of $-e^{-\phi}(\partial \chi/\partial t)T\Delta s$, where $\chi$
will increase as the star becomes more dense.

To estimate whether the contribution from these terms makes any
significant difference or we are just carrying coal to Newcastle, we
note that a transition from hadronic to quark matter may in a rough
approximation be seen as a transition between free Fermi gasses at
different densities--one of them with a bag constant in the energy
density, but this does not affect the entropy density. Assuming both
gasses are relativistic and degenerate the entropy per particle with
Boltzmann's constant set to one is \cite{landau80:a}
\begin{equation}
\sigma=\gamma_c\frac{(3\pi^2)^{\frac{2}{3}}}{3\hbar c}Tn^{-\frac{1}{3}}\simeq
\gamma_c 0.02 \frac{T}{\mathrm{MeV}}\left(\frac{n}{\mathrm{fm}^{-3}}
\right)^{-\frac{1}{3}}\;,
\end{equation}
where $\gamma_c$ is the color degeneracy, $n$ is the particle density
which for quark matter is three times the baryon density,
$n_\mathrm{QM}$, in the quark phase and in the hadronic phase is equal
to the baryon density there, $n_\mathrm{H}$. The resulting term in in
Eq. (\ref{balance}) is then on the order of
\begin{equation}
10^{33}\frac{T_9^2}{e^{\phi}}\left[\left(
  \frac{n_\mathrm{H}}{\mathrm{fm}^{-3}}\right)^{-\frac{1}{3}}-\left(
  \frac{3n_\mathrm{QM}}{\mathrm{fm}^{-3}}\right)^{-\frac{1}{3}}\right]
  \frac{\mathrm{d}N_q/\mathrm{d}t}{10^{57}/10^7\mathrm{yr}}\mathrm{erg}
  \mbox{ s}^{-1}\;,
\end{equation}
where $T_9=T/10^9$ K and $\mathrm{d}N_q/\mathrm{d}t$ is normalized to
a whole star converted on a timescale of $10^7$ years. For the quarks
$\gamma_c=3$ was canceled by a factor $1/3$ since it takes 3 quarks to
make a baryon. This is a rather small contribution unless the
temperature is high or the star is changing structure
fast. Furthermore the term is only positive if
$n_\mathrm{H}<3n_\mathrm{QM}$, but this condition is of course subject
to the specific assumptions we made here and would change in a more
detailed treatment of the EoS. It should also be noted that this
treatment accounts only for the latent heat from the deconfinement of
the quarks, and that there would be further contributions from the
subsequent weak reactions required to bring the matter into chemical
equilibrium by converting $d$-quarks to $s$-quarks. These reactions
would contribute to the $\epsilon_\mathrm{nuc}$ and $\epsilon_\nu$
terms in Eq. (\ref{balance}) and must be treated similarly to the
rotochemical heating discussed by \cite{fernandez05:a}. As shown by
\cite{kaminker06:a} it is furthermore very important to know exactly
where in the star a heating source is located, since sources located
below the outer crust of the star tend to contribute only to the
neutrino luminosity and make little difference to the surface
temperature.  In spite of such reservations one might still
optimistically expect to find signals from the latent heat of a quark
matter phase transition either early in a star's thermal history, when
it is hot and spinning down rapidly, or from later episodes of rapid
structure change. Such rapid structure change could result from
glitches, which may be associated with the buildup and release of
stress in a crystalline mixed quark hadron phase \cite{glen95:a}, or
the appearance of quark matter in a neutron star core, which softens
the equations of state and allows the star to change structure rapidly
possibly even through a core quake if a metastable state can be
accessed \cite{zdunik06:a}. Future work will explore these
possibilities.

\goodbreak
\section{Summary}

Obviously, our view of the interior composition of pulsars, which
contain matter in one of the densest forms found in the Universe, has
changed dramatically since their first discovery some 40 years ago. It
has also become clear during that time period that all the ambient
conditions that characterize pulsars tend to the extreme as well,
rendering pulsars to almost ideal astrophysical laboratories for a
broad range of physical studies. Owing to the unprecedented wealth of
high-quality data on pulsars provided by radio telescopes, X-ray
satellites--and soon the latest generation of gravitational-wave
detectors--it seems within reach to decipher the inner workings of
these enigmatic objects, and to explore the phase diagram of cold and
ultra-dense hadronic matter from astrophysics.



\section*{Acknowledgments}

The research of F.\ Weber is supported by the National Science
Foundation under Grant PHY-0457329, and by the Research Corporation.
 


\end{document}